\documentclass{ecai} 

\usepackage{latexsym}
\usepackage{amssymb}
\usepackage{amsmath}
\usepackage{amsthm}
\usepackage{booktabs}
\usepackage{enumitem}
\usepackage{graphicx}
\usepackage{multirow}
\usepackage{color}
\usepackage{ulem}

\newcommand{\BibTeX}{B\kern-.05em{\sc i\kern-.025em b}\kern-.08em\TeX}

\begin{document}


\begin{frontmatter}


\paperid{123} 


\title{LexSemBridge: Fine-Grained Dense Representation Enhancement through Token-Aware Embedding Augmentation}



\author[A]{\fnms{Shaoxiong}~\snm{Zhan}\footnote{Equal contribution.}}
\author[A,B]{\fnms{Hai}~\snm{Lin}\footnotemark}
\author[A,B]{\fnms{Hongming}~\snm{Tan}} 
\author[A]{\fnms{Xiaodong}~\snm{Cai}}
\author[A,B]{\fnms{Hai-Tao}~\snm{Zheng}\thanks{Corresponding Author. Email: zheng.haitao@sz.tsinghua.edu.cn.}}
\author[C]{\fnms{Xin}~\snm{Su}}
\author[C]{\fnms{Zifei }~\snm{Shan}}
\author[A]{\fnms{Ruitong }~\snm{Liu}}
\author[D]{\fnms{Hong-Gee }~\snm{Kim}}

\address[A]{Shenzhen International Graduate School, Tsinghua University, Shenzhen, China}
\address[B]{Pengcheng Laboratory, Shenzhen, China}
\address[C]{Tencent Company, Shenzhen, China}
\address[D]{Seoul National University, South Korea}


\begin{abstract}

As queries in retrieval-augmented generation (RAG) pipelines powered by large language models (LLMs) become increasingly complex and diverse, dense retrieval models have demonstrated strong performance in semantic matching. Nevertheless, they often struggle with fine-grained retrieval tasks, where precise keyword alignment and span-level localization are required, even in cases with high lexical overlap that would intuitively suggest easier retrieval. To systematically evaluate this limitation, we introduce two targeted tasks, keyword retrieval and part-of-passage retrieval, designed to simulate practical fine-grained scenarios. Motivated by these observations, we propose LexSemBridge, a unified framework that enhances dense query representations through fine-grained, input-aware vector modulation. LexSemBridge constructs latent enhancement vectors from input tokens using three paradigms: Statistical (SLR), Learned (LLR), and Contextual (CLR), and integrates them with dense embeddings via element-wise interaction. Theoretically, we show that this modulation preserves the semantic direction while selectively amplifying discriminative dimensions. LexSemBridge operates as a plug-in without modifying the backbone encoder and naturally extends to both text and vision modalities. Extensive experiments across semantic and fine-grained retrieval tasks validate the effectiveness and generality of our approach. All code and models are publicly available at \url{https://github.com/Jasaxion/LexSemBridge/}

\end{abstract}

\end{frontmatter}

\section{Introduction}

Large language models (LLMs) have significantly advanced the capabilities of information retrieval systems, particularly within the retrieval-augmented generation (RAG) paradigm, where dense representations serve as a core mechanism to access relevant knowledge~\cite{lirefine, lione, wang2024large, li2025towards, xu2025let}. While dense retrieval models are effective at capturing coarse-grained semantic similarity, which focuses on retrieving passages broadly related to the query topic, they are less suitable for fine-grained retrieval scenarios that require precise lexical or structural correspondence~\cite{karpukhin2020dense, luan2021sparse}. Fine-grained retrieval emphasizes matching specific terms, entities, or short spans within documents, in contrast to coarse-grained retrieval, which targets semantically related but potentially diffuse content.

Such fine-grained information needs frequently arise in practical applications, including open-domain question answering, fact verification, and document-grounded generation, where retrieving exact keywords or localized evidence segments is critical~\cite{lin2024irsc, yan2024atomic}. Although existing benchmarks such as MTEB~\cite{muennighoff2023mteb} have standardized evaluations of retrieval models, they predominantly assess coarse-grained semantic matching, providing limited insights into models' capabilities under fine-grained retrieval requirements.

To facilitate a more targeted evaluation, we introduce two fine-grained retrieval tasks: keyword retrieval, where the query is a salient term extracted from the ground-truth passage, and part of passage (P-o-P) retrieval, where the query corresponds to a short contiguous span. These tasks are designed to simulate practical retrieval demands that prioritize precise alignment over broad semantic relevance.

Our analysis reveals that while dense retrievers achieve strong performance on semantic benchmarks, they exhibit a substantial performance decline on fine-grained tasks. In particular, for the part of passage retrieval setting, where queries are directly extracted spans from the original passages, dense models frequently fail to retrieve the corresponding source segments, despite the high lexical overlap. This unexpected failure highlights the limitations of dense encoders, whose holistic optimization objectives prioritize global semantic abstraction while often overlooking token-level or localized cues critical for fine-grained retrieval~\cite{karpukhin2020dense, lee2019latent, luan2021sparse}.

In this work, we propose \textbf{LexSemBridge} (Lexical-Semantic Bridge), a unified framework for dense representation enhancement that strengthens fine-grained retrieval capabilities without sacrificing semantic expressiveness. LexSemBridge constructs token-aware enhancement vectors based on the input, which are fused into the original dense representations through element-wise modulation. We instantiate this framework using three complementary paradigms, namely Statistical Lexical Representation (SLR), Learned Lexical Representation (LLR), and Contextual Lexical Representation (CLR) in Sec.~\ref{sec:lrc}, each aiming to capture different aspects of token-level importance. Notably, LexSemBridge can be incorporated into existing retrieval models without requiring modifications to the backbone encoder, making it a lightweight and broadly applicable enhancement. Beyond textual retrieval, we demonstrate that LexSemBridge naturally generalizes to vision-based retrieval tasks by treating visual patches as token analogs~\cite{baobeit}, allowing a unified paradigm to improve fine-grained dense representations across modalities.

Our contributions are summarized as follows. 1) We propose LexSemBridge, a modular and lightweight framework that enhances the fine-grained capabilities of dense representations while preserving their original semantic structure. 2) To better characterize the limitations of dense models in capturing fine-grained signals, we introduce two diagnostic retrieval tasks, keyword retrieval and part of passage retrieval, which are designed to reflect retrieval needs commonly encountered in large-scale retrieval augmented generation (RAG) systems. 3) We show that LexSemBridge generalizes beyond textual retrieval, achieving consistent improvements in vision-based tasks and demonstrating its applicability across modalities.

\section{Related Work}

With the rapid development of pre-trained Transformer models (PLMs) \cite{vaswani2017attention}, dense representations have become a central component in modern information access systems, driven by large-scale pre-trained encoders that project queries and documents into a shared semantic vector space~\cite{yates2021pretrained, karpukhin2020dense, zhao2024dense}. Early models such as BERT and its derivatives~\cite{devlin2019bert, sanh2019distilbert, song2020mpnet, Reimers2019SentenceBERT} typically adopt a dual encoder architecture and generate a single embedding per input, allowing efficient retrieval via vector similarity. More recent efforts, including BGE-M3~\cite{chen2024bge}, introduce richer representational structures to capture diverse semantic aspects within input sequences. Similarly, GTE~\cite{zhang2024mgte} improves long-context and multilingual representations through hierarchical contrastive pre-training, while Snowflake~\cite{merrick2024embedding} increases embedding quality through large-scale contrastive training with stratified sampling. These advancements have led to consistent improvements across benchmark datasets, as reflected in leaderboards such as MTEB~\cite{muennighoff2023mteb}. However, existing dense encoders often encounter limitations in fine-grained retrieval scenarios that require high token-level resolution, such as keyword-based or segment-level matching~\cite{lee2019latent, lin2024irsc}. These challenges are largely attributed to coarse-grained pooling operations and insufficient modeling of localized lexical semantics~\cite{karpukhin2020dense, lee2019latent, luan2021sparse}. This has motivated efforts to improve dense representations at the embedding level in a model-agnostic manner, while maintaining compatibility with standard retrieval architectures.

Previous work has explored various approaches to enhance the overall performance of retrieval systems, addressing multiple aspects of representation learning. ColBERT~\cite{khattab2020colbert} introduces a late interaction retrieval paradigm that preserves token-level embeddings through intermediate-layer multi-vector representations, requiring complex indexing structures and computationally intensive interaction mechanisms, which ultimately modifies the conventional retrieval framework. SPARTA~\cite{zhao2021sparta} utilizes tokenizer-derived outputs for queries and dense BERT embeddings for passages, computing relevance based on the similarity between sparse lexical tokens and dense semantic vectors. SPLADE~\cite{formal2021splade}, on the other hand, produces sparse lexical representations from dense encoder outputs using expansion and sparsification techniques. While these approaches offer valuable insights into latent structure modeling, they often rely on sparse token-level representations or require substantial architectural modifications and non-trivial infrastructure. In contrast, our work focuses on enhancing the latent semantics of dense representations through a lightweight, token-aware augmentation mechanism. The proposed method operates entirely within the dense representation space and can be readily integrated into standard encoder architectures without modifying the retrieval framework. Moreover, this formulation is modality-agnostic and may be extended to non-textual inputs with discrete token structures, as in vision transformers that represent images as patch tokens~\cite{baobeit}.

\section{Methodology}

\subsection{Preliminaries: Dense Retrieval and Vector-Based Representation}

Dense retrieval has become a key paradigm in modern information access, especially in RAG systems and language model applications~\cite{izacard2023atlas}. Unlike sparse methods such as BM25 that depend on exact token matches, dense retrieval encodes queries and documents into continuous embeddings, enabling semantic similarity through learned representations.

Formally, given a corpus of passages \( \mathcal{P} = \{p_1, ..., p_n\} \) and a query \( q \), dense retrievers employ a dual encoder framework consisting of a query encoder \( \text{Enc}_Q \) and a passage encoder \( \text{Enc}_P \). Each input is mapped to a dense vector through contextualized pooling:
\begin{equation}
\label{eq:embedding_eq}
\begin{split}
\mathbf{e}_q = \text{Pooling}(\text{Enc}_Q(q)) \\
\mathbf{e}_p = \text{Pooling}(\text{Enc}_P(p))
\end{split}
\end{equation}

Common pooling strategies include mean pooling or selecting ``[CLS]'' token embedding. Relevance between a query and passage is typically measured using cosine similarity:
\begin{equation}
\label{eq:cosim}
\text{sim}_{\text{dense}}(q, p) = \frac{\langle \mathbf{e}_q, \mathbf{e}_p \rangle}{\|\mathbf{e}_q\| \cdot \|\mathbf{e}_p\|}
\end{equation}

This framework supports scalable, end-to-end trainable retrieval. However, its focus on global semantic alignment may overlook fine-grained lexical signals. To address this, we enhance dense representations with task-aware lexical modulation while preserving the original vector structure and similarity computation, forming the basis of our unified vector enhancement framework.

\subsection{Preliminaries: Masked Language Modeling and Contrastive Learning}
\label{sec:prelim_mlm_contrastive}

\textbf{Masked Language Modeling (MLM).}  
MLM is a standard pre-training objective for transformer-based language models. Given an input sequence \( x = (x_1, \ldots, x_n) \), a subset of tokens is replaced with a special ``[MASK]'' symbol. The model learns to recover masked tokens from their context. Let \( h \in \mathbb{R}^d \) be the hidden state in a masked position and \( \mathcal{V} \) be the vocabulary. The output distribution over tokens is:
\begin{equation}
\label{eq:mlm_head}
P(x_i | x_{\setminus i}) = \frac{\exp(v_i^\top g(h))}{\sum_{j \in \mathcal{V}} \exp(v_j^\top g(h))},
\end{equation}
where \( g(h) = \text{LayerNorm}(Wh + b) \) and \( v_i \in \mathbb{R}^d \) is the output embedding for token \( i \).

\textbf{Contrastive Learning.}  
To learn discriminative dense representations, contrastive learning is widely used in dual encoder retrieval models. Given a batch of query-passage pairs \( \{(q_i, p_i)\}_{i=1}^B \), each query is trained to be most similar to its corresponding passage and dissimilar to negatives. The loss is defined as follows:
\begin{equation}
\mathcal{L}_{\text{contrast}} = -\frac{1}{B} \sum_{i=1}^{B} \log \frac{\exp(\text{Sim}(\mathbf{e}_{q_i}, \mathbf{e}_{p_i}) / \tau)}{\sum_{j=1}^{G} \exp(\text{Sim}(\mathbf{e}_{q_i}, \mathbf{e}_{p_j}) / \tau)},
\end{equation}
where \( \text{Sim}(\cdot, \cdot) \) denotes cosine similarity, \( \tau \) is a temperature parameter, and \( G \) is the number of passages (one positive and \( G-1 \) negatives). These two objectives form the theoretical basis for many neural language models and retrieval architectures.

\subsection{Fine-Grained Retrieval Task Formulation}

To evaluate the capability of dense retrieval models across varying levels of semantic granularity, we consider four retrieval settings unified under a cosine similarity-based dual encoder framework. Each task is defined as a top-\(k\) nearest neighbor retrieval problem in the dense embedding space.

\begin{figure}[h]
  \centering
  \includegraphics[width=0.92\linewidth]{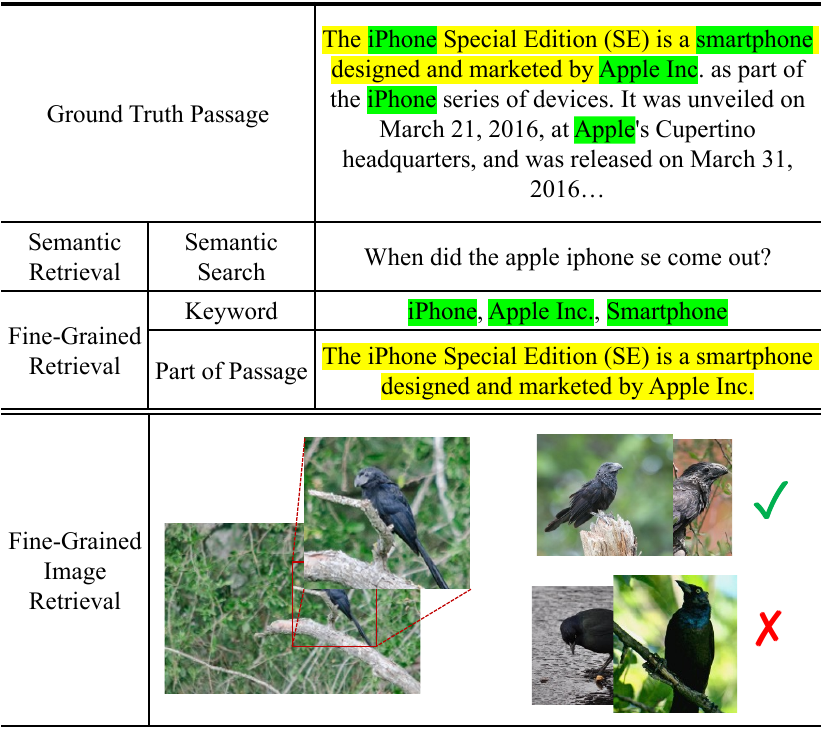}
  \caption{Four fine-grained retrieval tasks spanning semantic, lexical, and visual levels, reflecting potential real-world query scenarios in large-scale RAG-based retrieval systems.}
  \label{Example_tasks}
\end{figure}

Let \( \mathcal{P} = \{p_1, ..., p_n\} \) be a candidate set of passages or images, and \( q \) a given query (text or visual). The model ranks items in \( \mathcal{P} \) by computing the cosine similarity in Eq.~\eqref{eq:cosim}. Figure~\ref{Example_tasks} illustrates example scenarios corresponding to the four retrieval tasks.

We examine four tasks:
\begin{itemize}
    \item \textbf{Semantic Search.}  
    The goal is to retrieve passages topically or semantically relevant to a natural language query. This task is the default benchmark scenario in datasets such as BEIR \cite{thakur2021beir} and MTEB \cite{muennighoff2023mteb}, and it evaluates global semantic understanding in dense retrievers.

    \item \textbf{Keyword Matching.}  
    Given a query composed of one or more keywords \( q = \{k_1, \ldots, k_m\} \subset \mathcal{V} \), the goal is to retrieve the original passage \( p \in \mathcal{P} \) containing them. This task emphasizes sparse lexical cues rather than a full semantic context, which poses a greater challenge for dense encoders.

    \item \textbf{Part of Passage (P-o-P) Retrieval.}  
    Let a passage \( p \in \mathcal{P} \) contain a contiguous text span \( s \subset p \). The query is \( q = s \), and the objective is to retrieve the full passage \( p \). This task simulates span-level understanding, focusing on whether local textual context is adequately preserved in the global embedding.

    \item \textbf{Fine-Grained Image Retrieval.}  
    In the vision domain, we evaluate retrieval using full image queries over an image corpus. Images in this setting often belong to the same category or subclass (e.g., bird species, car models), with only subtle visual differences. As such, this task inherently possesses a fine-grained nature and serves as the image-side counterpart to textual fine-grained retrieval.
\end{itemize}

The latter three tasks serve as diagnostic probes for assessing the preservation of localized information in dense representations. Although there are related formulations \cite{miao2020keyword, lin2024irsc}, we adopt a unified evaluation framework with cosine-based dual encoders in both textual and visual modalities.

\subsection{Lexical Representation Construction}
\label{sec:lrc}


To construct enhancement vectors that capture fine-grained lexical signals, we analyze the original input sequence of length \( L \) from multiple perspectives and extract a vocabulary-level importance vector \( \mathbf{w} \in \mathbb{R}^{|\mathcal{V}|} \), where \( \mathcal{V} \) is the encoder’s vocabulary and the hidden dimension is \( d \). To ensure numerical stability and avoid large disparities in importance values, we apply log1p normalization to \( \mathbf{w} \). We explore three paradigms for deriving \( \mathbf{w} \), differing in whether they rely on statistical frequency, encoder-derived contextual signals, or model-internal predictions.

\paragraph{Statistical Lexical Representation (SLR).}
SLR derives a binary indicator vector from token presence in the input sequence. Given an input sequence \( T = [t_1, ..., t_L] \), we define:
\begin{equation}
\label{eq:slr}
  w_{\text{SLR}}(t) = \log(1+\max_{j=1...L} \mathbf{1}[t = t_j]), \quad t \in \mathcal{V}
\end{equation}
This formulation produces a sparse, interpretable vector that reflects which vocabulary items occur in the input.

\paragraph{Learned Lexical Representation (LLR).}
LLR uses encoder hidden states to compute contextualized lexical relevance. Given token embeddings \( H = [h_1, ..., h_L] \in \mathbb{R}^{L \times d} \), we compute:
\begin{equation}
\label{eq:llr}
  w_{\text{LLR}}(t) = \log(1+\text{ReLU}(h_j U + b)), \quad t \in \mathcal{V}
\end{equation}
where \( U \in \mathbb{R}^{d \times |\mathcal{V}|} \) is a learned projection matrix and \( b \in \mathbb{R}^{|\mathcal{V}|} \) is a bias term. The ReLU ensures non-negativity, while the logarithmic transformation compresses the activation range.


\paragraph{Contextual Lexical Representation (CLR).}
CLR leverages the MLM prediction head of the pre-trained encoder in Eq.~\eqref{eq:mlm_head} to extract token-level salience from the ``[CLS]'' embedding. Let \( h_0 \) denote the ``[CLS]'' hidden state. Then:
\begin{equation}
\label{eq:clr}
  w_{\text{CLR}}(t) = \log(1+\text{MLM-Head}(h_0)_t), \quad t \in \mathcal{V}
\end{equation}
This representation captures the model's global contextual expectations over the vocabulary, based on the entire input.  
Since MLM-Head produces a softmax-normalized probability distribution, \( w_{\text{CLR}}(t) \) corresponds to the log-probability rather than the original logits.

\paragraph{From Importance Vector to Enhancement.}
Each strategy yields a vocabulary-level importance vector \( \mathbf{w} \), which is projected by a trainable matrix \( \mathbf{W} \in \mathbb{R}^{d \times |\mathcal{V}|} \). We apply the softmax function to normalize the resulting scores:
\begin{equation}
  \label{eq:qlex_ref}
  \mathbf{q}_{\text{lex}} = \texttt{softmax}(\mathbf{W} \cdot \mathbf{w})
\end{equation}
This produces a distribution emphasizing the relative importance of each lexical feature. We find this yields more stable results in practice. The enhancement vector \( \mathbf{q}_{\text{lex}} \in \mathbb{R}^d \) is then fused with the original query vector via element-wise multiplication (Sec.~\ref{sec:method_detail}), enabling consistent integration of lexical information.

\subsection{Semantic Preservation and Fine-Grained Enhancement}
\label{sec:method_detail}

Let \( \mathbf{q}_{\text{dense}} \in \mathbb{R}^{d} \) denote the original dense query vector. We compute an enhancement vector \( \mathbf{q}_{\text{lex}} \in \mathbb{R}^{d} \) by projecting a lexical distribution onto a learned matrix as defined in Eq.~\eqref{eq:qlex_ref}. The final query representation \( \mathbf{q}_{\text{out}} \) is obtained via element-wise multiplication:
\begin{equation}
\label{eq:qoutexp}
\mathbf{q}_{\text{out}} = \mathbf{q}_{\text{dense}} \odot \mathbf{q}_{\text{lex}}.
\end{equation}

We base our theoretical framework on the assumption that embedding dimensions inherently differ in their sensitivity to semantic and lexical information, a property consistently observed in transformer-based representations. Specifically, some dimensions predominantly encode high-level semantic concepts, while others focus on fine-grained lexical details. We formally define \( \mathcal{S}_{\text{sem}} \subseteq \{1, \dots, d\} \) as the set of dimensions capturing semantic content, and \( \mathcal{S}_{\text{lex}} \subseteq \{1, \dots, d\} \) as the set capturing lexical-specific features.

To illustrate, consider the query ``symptoms and treatment of diabetes.'' Dimensions in \( \mathcal{S}_{\text{sem}} \) encode general themes such as ``health conditions'' or ``medical interventions,'' which help retrieve broadly relevant documents. In contrast, dimensions in \( \mathcal{S}_{\text{lex}} \) emphasize specific terms like ``symptoms,'' ``treatment,'' and ``diabetes,'' enabling the retrieval of documents precisely addressing the detailed aspects mentioned in the query.

Thus, the enhancement vector \( \mathbf{q}_{\text{lex}} \) selectively scales lexical-sensitive dimensions more strongly than semantic ones, represented as:
\begin{equation}
\mathbf{q}_{\text{lex}, i} \approx c, \quad \forall i \in \mathcal{S}_{\text{sem}}, \quad \mathbf{q}_{\text{lex}, j} \gg c, \quad \forall j \in \mathcal{S}_{\text{lex}},
\end{equation}
where \( c \) is a positive baseline constant. Specifically, \( c \) reflects the approximate mean scaling factor over semantic-sensitive dimensions, resulting from the aggregation of lexical importance values across the vocabulary. Since the majority of vocabulary tokens exhibit relatively low and smoothly distributed lexical importance, their projected contributions induce near-uniform scaling over \( \mathcal{S}_{\text{sem}} \). This design ensures that the semantic alignment of the query embedding remains largely unaffected during the enhancement process.

Empirical validation of this hypothesis involves:
\begin{itemize}
\item Evaluating different LexSemBridge strategies (SLR, LLR, CLR) in Sec.~\ref{sec:main_result}, demonstrating consistent improvements on fine-grained retrieval tasks without degrading semantic retrieval performance;
\item Conducting ablation studies in Sec.~\ref{sec:ablation_qencoer} and~\ref{exp:ablation_lsb} to isolate the contributions of enhancement vectors and fusion mechanisms, validating the necessity of the proposed design;
\item Extending LexSemBridge to image retrieval tasks in Sec.~\ref{sec:ad_to_image}, confirming the modality-general applicability of the selective enhancement mechanism.
\end{itemize}

Through targeted evaluations, LexSemBridge demonstrates improved fine-grained retrieval sensitivity without compromising overall semantic performance. This is achieved by reweighting the dimensions of \( \mathbf{q}_{\text{dense}} \) to selectively emphasize fine-grained features while preserving global semantic alignment, leading to stable optimization and enhanced retrieval accuracy.

\subsection{Extension to Vision: Patch-Aware Modulation}
Our method directly applies to vision encoders by treating image patches analogously to tokens. For example, models like BEiT~\cite{baobeit} tokenize image patches and encode them as embeddings. A patch-level relevance vector is computed and projected to the same embedding space:
\begin{equation}
\begin{split}
\mathbf{q}_{\text{lex}} = \texttt{softmax}(\mathbf{W}_{\text{patch}} \cdot \mathbf{r}_{\text{patch}})
\end{split}
\end{equation}
where \( \mathbf{r}_{\text{patch}} \in \mathbb{R}^{N} \) represents patch importance. This enables patch-aware modulation for visual queries. The fusion remains unchanged:
\begin{equation}
\begin{split}
\mathbf{q}_{\text{Image\_out}} = \mathbf{q}_{\text{Image\_dense}} \odot \mathbf{q}_{\text{lex}}
\end{split}
\end{equation}

This adaptation bridges lexical-semantic fusion with visual-semantic grounding, forming a unified and modality-agnostic representation framework.

\subsection{Lexical-Semantic Integration}


To incorporate lexical priors into dense retrieval, we modulate the query representation via a lightweight lexical-semantic bridge. Given the original query vector \( \mathbf{q}_{\text{dense}} \), we compute a vocabulary-level importance vector \( \mathbf{w} \) using one of three strategies. The Statistical Lexical Representation (SLR, Eq.~\eqref{eq:slr}) identifies tokens explicitly present in the query. The Learned Lexical Representation (LLR, Eq.~\eqref{eq:llr}) assigns relevance scores to semantically related terms based on encoder hidden states. The Contextual Lexical Representation (CLR, Eq.~\eqref{eq:clr}) estimates a full-vocabulary importance distribution using the masked language modeling head. The vector \( \mathbf{w} \) is then projected via a trainable linear transformation (Linear Alignment Layer) and normalized with softmax to obtain the enhancement vector \( \mathbf{q}_{\text{lex}} \) (Eq.~\eqref{eq:qlex_ref}).

The final query representation \( \mathbf{q}_{\text{out}} \) is obtained via element-wise modulation with \( \mathbf{q}_{\text{lex}} \), as previously described in Sec.~\ref{sec:method_detail}. This operation reallocates representational focus toward semantically aligned lexical dimensions, improving fine-grained discrimination.

\begin{figure}[h]
  \centering
  \includegraphics[width=1.0\linewidth]{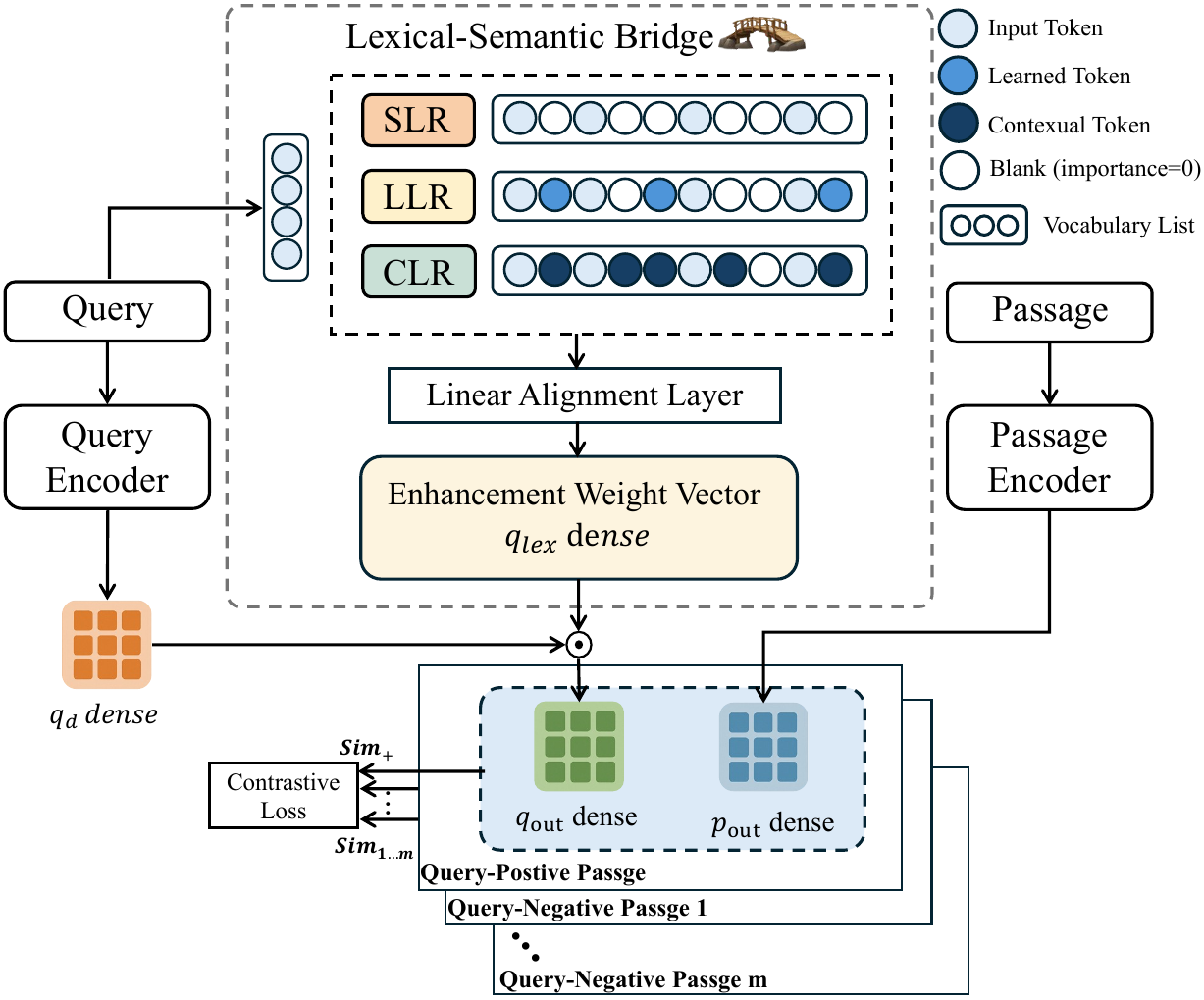}
  \caption{Overview of the complete framework architecture with training process. The Lexical-Semantic Bridge module can be incorporated as a plugin into the training pipeline of other encoding models.}
  \label{LexSem_bridge}
\end{figure}

On the passage side, we retain the original dense representation \( \mathbf{p}_{\text{dense}} \) without lexical modulation. This design reflects the retrieval-specific intuition that lexical cues are primarily query-driven, and enhancing only the query side is sufficient to improve matching performance while avoiding additional computational overhead. We further validate this asymmetry through an ablation study in Section~\ref{sec:ablation_qencoer}, which analyzes the effects of applying enhancement to either or both sides. As shown in Figure~\ref{LexSem_bridge}, the framework is encoder-agnostic and integrates seamlessly with standard contrastive training pipelines.

By enriching the query-side representation while maintaining architectural simplicity, our method enhances fine-grained expressiveness without disrupting the dense retrieval objective.

\section{Experiments}
\label{sec:exp}

\subsection{Base Models}
\label{sec:base_model}


To demonstrate the generalizability of our framework, we conduct experiments on multiple encoder backbones. DistilBERT and MPNet are used as lightweight models for ablation, while GTE and Snowflake-Arctic-Embed-L serve as stronger baselines to evaluate compatibility with state-of-the-art retrievers. For image retrieval, we use BEiT-base to validate performance. All base models are trained with the same supervision and optimization pipeline, ensuring fair comparison.

\subsection{Datasets and Evaluation Metrics}
\label{datasets_exp}


\paragraph{Training Data.} We utilize the All-NLI\footnote{\url{https://huggingface.co/datasets/sentence-transformers/all-nli}} dataset as the basis for model fine-tuning, enhanced through Hard Negative Sampling techniques \cite{Wang2022SimlmPB, Lee2021ApproximateNN}. The dataset has been curated to fit the contrastive learning paradigm, providing entailment and contradiction pairs as positive and negative examples, respectively, and exhibits low overlap with widely-used datasets like MS MARCO. For sampling implementation, we employ the BGE-base-en-v1.5 model as the retrieval engine and randomly select 15 samples from positions 20 to 200 in the retrieval results as negative samples. This approach enhances both the quality of training data and training efficiency.


\paragraph{Evaluation Data.} We employ three widely-used public datasets: HotpotQA~\cite{yang2018hotpotqa}, FEVER~\cite{thorne2018fever}, and Natural Questions (NQ)~\cite{kwiatkowski2019natural}. The Semantic Search task is sampled from BEIR/MTEB benchmarks, while the Keyword and Part of Passage tasks are not included in BEIR/MTEB and are introduced to evaluate fine-grained retrieval capabilities. We extract 5,000 entries from both HotpotQA and NQ, and use the FEVER test set containing 1,499 entries as evaluation samples. Each dataset is used to construct three retrieval tasks:
\begin{itemize}
    \item \textbf{Semantic Search}: The model is tasked with retrieving passages that are semantically related to the query. It must retrieve passages that align with the semantic content of the query.
    \item \textbf{Keyword Matching}: Queries consist of 3 to 8 keywords extracted from the original passage using a large language model. The model’s task is to retrieve passages containing these keywords.
    \item \textbf{Part of Passage (P-o-P)}: We extract text spans of 16, 32, 64, 128, and 256 words from the original passages, and the model is tasked with retrieving the full passage containing the span.
\end{itemize}

Additionally, for the image retrieval task, we conduct transfer experiments using the complete training and test sets from CUB-200~\cite{wah2011caltech} and Stanford Cars~\cite{krause20133d}. Given a query image, the model's task is to retrieve images that belong to the same category as the query.

\paragraph{Metrics.} We report normalized discounted cumulative gain (nDCG) as the primary metric using the BEIR~\cite{thakur2021beir} evaluation toolkit. nDCG measures ranking quality with position sensitivity, and is widely adopted in retrieval literature. Full metrics including MRR and Recall@K are included in the supplementary material.

\subsection{Finetune Strategy}
\label{sec:finetune_strategy}

All models are fine-tuned using contrastive learning with a temperature-scaled cross-entropy loss, as defined in Section~\ref{sec:prelim_mlm_contrastive}. Training is conducted on 8$\times$A100 GPUs with a batch size of 64 for 10 epochs. We use a learning rate of \(1 \times 10^{-5}\), mixed precision (FP16), and a temperature scaling factor of 0.02. The framework supports the following configurations:
\begin{itemize}
    \item \textbf{Baseline}: A dual encoder trained on semantic matching only, without lexical enhancement.
    \item \textbf{SLR-enhanced}: Incorporates statistical lexical cues from token presence (Eq.~\eqref{eq:slr}).
    \item \textbf{LLR-enhanced}: Computes learned lexical importance using hidden states (Eq.~\eqref{eq:llr}).
    \item \textbf{CLR-enhanced}: Utilizes contextual token salience from the MLM prediction head (Eq.~\eqref{eq:clr}).
\end{itemize}
All models are trained with the same supervision and optimization pipeline, ensuring fair comparison. The training setup includes a query max length of 64 tokens, passage max length of 256 tokens, logging every 10 steps, and saving checkpoints every 5000 steps.

\subsection{Analysis of Retrieval Performance with Lexical-Semantic Bridge}
\label{sec:main_result}

\begin{table*}[h]
\centering
\caption{Comparative Evaluation of Lexical-Semantic Bridge (SLR/LLR/CLR) and Neural Baselines: nDCG@1($\times$100) on Semantic Search (Query), Keyword Search (Keyword), and Part of Passage (P-o-P) 16/32/64 Tasks Across Public Datasets. The best performance is marked in \textbf{bold}, and the second-best performance is \uline{underlined}.}
\label{tab:validate_result}
\resizebox{0.96\textwidth}{!}{%
\begin{tabular}{l|lllll|lllll|lllll}
\hline
\multirow{3}{*}{Model} & \multicolumn{5}{c|}{HotpotQA} & \multicolumn{5}{c|}{FEVER} & \multicolumn{5}{c}{NQ} \\
 & \multirow{2}{*}{Query} & \multirow{2}{*}{Keyword} & \multicolumn{3}{c|}{P-o-P} & \multirow{2}{*}{Query} & \multirow{2}{*}{Keyword} & \multicolumn{3}{c|}{P-o-P} & \multirow{2}{*}{Query} & \multirow{2}{*}{Keyword} & \multicolumn{3}{c}{P-o-P} \\
 &  &  & \multicolumn{1}{c}{16} & \multicolumn{1}{c}{32} & \multicolumn{1}{c|}{64} &  &  & \multicolumn{1}{c}{16} & \multicolumn{1}{c}{32} & \multicolumn{1}{c|}{64} &  &  & \multicolumn{1}{c}{16} & \multicolumn{1}{c}{32} & \multicolumn{1}{c}{64} \\ \hline
DistilBERT-base $*$ & 22.77 & 58.72 & 80.30 & 96.30 & \textbf{99.34} & 31.83 & 50.03 & 50.23 & 70.51 & {\uline{88.13}} & 13.64 & 39.12 & 71.88 & 92.00 & {\uline{98.54}} \\
\hspace{1em} +Baseline & 35.15 & 66.54 & 80.26 & 96.86 & 99.26 & 49.78 & 56.97 & 46.16 & 69.91 & 83.99 & 19.70 & 49.91 & 70.36 & 91.30 & 98.06 \\
\hspace{1em} +SLR & 36.63 & 69.16 & 81.80 & 97.02 & {\uline{99.28}} & 53.03 & 59.69 & 47.97 & 70.71 & 84.06 & 20.46 & 52.92 & 72.22 & 91.60 & 98.10 \\
\hspace{1em} +LLR & {\uline{38.61}} & {\uline{76.31}} & {\uline{85.20}} & {\uline{97.40}} & {\uline{99.28}} & {\uline{62.08}} & {\uline{66.01}} & {\uline{53.64}} & {\uline{73.58}} & 86.26 & {\uline{25.76}} & {\uline{63.47}} & {\uline{78.12}} & {\uline{93.06}} & 98.50 \\
\hspace{1em} +CLR & \textbf{44.06} & \textbf{79.03} & \textbf{86.20} & \textbf{97.64} & \textbf{99.34} & \textbf{68.51} & \textbf{72.81} & \textbf{58.31} & \textbf{77.39} & \textbf{90.19} & \textbf{28.79} & \textbf{67.78} & \textbf{79.20} & \textbf{93.98} & \textbf{98.92} \\ \hline \hline
MPNet-base $*$ & 6.44 & 8.23 & 50.36 & 86.00 & 98.50 & 15.35 & 12.71 & 22.75 & 50.97 & 75.05 & 5.30 & 5.82 & 41.40 & 77.46 & 96.08 \\
\hspace{1em} +Baseline & 40.10 & 72.84 & 85.62 & 98.16 & {\uline{99.44}} & 79.60 & 75.12 & 61.91 & 83.59 & 94.13 & 27.27 & 56.67 & 79.02 & 95.04 & 99.04 \\
\hspace{1em} +SLR & 42.08 & 73.56 & 85.78 & {\uline{98.18}} & \textbf{99.46} & 80.63 & 74.92 & 62.91 & 84.06 & \textbf{94.33} & {\uline{29.55}} & 58.92 & 79.36 & {\uline{95.14}} & \textbf{99.14} \\
\hspace{1em} +LLR & {\uline{43.56}} & {\uline{74.89}} & {\uline{86.32}} & \textbf{98.26} & {\uline{99.44}} & {\uline{81.10}} & {\uline{75.53}} & \textbf{63.51} & {\uline{84.26}} & {\uline{94.26}} & 28.79 & {\uline{59.76}} & {\uline{79.74}} & 95.12 & 99.04 \\
\hspace{1em} +CLR & \textbf{44.55} & \textbf{78.31} & \textbf{87.52} & 98.16 & \textbf{99.46} & \textbf{82.79} & \textbf{77.57} & {\uline{63.38}} & \textbf{84.32} & 94.20 & \textbf{31.82} & \textbf{65.20} & \textbf{81.88} & \textbf{95.58} & {\uline{99.12}} \\ \hline
\multicolumn{4}{l}{\small $*$ denotes the original model.}\\ 
\end{tabular}%
}
\end{table*}

\begin{table*}[h]
\centering
\caption{Comparative Evaluation of Lexical-Semantic Bridge and State-of-the-Art Embedding Models: nDCG@1($\times$100) on Semantic Search (Query), Keyword Search (Keyword), and Part of Passage (P-o-P) 16 Tasks Across Public Datasets}
\label{tab:plugintosotamodel}
\resizebox{0.85\textwidth}{!}{%
\begin{tabular}{l|lll|lll|lll}
\hline
\multirow{3}{*}{Model} & \multicolumn{3}{c|}{HotpotQA} & \multicolumn{3}{c|}{FEVER} & \multicolumn{3}{c}{NQ} \\
 & \multicolumn{1}{c}{\multirow{2}{*}{Query}} & \multicolumn{1}{c}{\multirow{2}{*}{Keyword}} & \multicolumn{1}{c|}{\multirow{2}{*}{P-o-P-16}} & \multicolumn{1}{c}{\multirow{2}{*}{Query}} & \multicolumn{1}{c}{\multirow{2}{*}{Keyword}} & \multicolumn{1}{c|}{\multirow{2}{*}{P-o-P-16}} & \multicolumn{1}{c}{\multirow{2}{*}{Query}} & \multicolumn{1}{c}{\multirow{2}{*}{Keyword}} & \multicolumn{1}{c}{\multirow{2}{*}{P-o-P-16}} \\
 & \multicolumn{1}{c}{} & \multicolumn{1}{c}{} & \multicolumn{1}{c|}{} & \multicolumn{1}{c}{} & \multicolumn{1}{c}{} & \multicolumn{1}{c|}{} & \multicolumn{1}{c}{} & \multicolumn{1}{c}{} & \multicolumn{1}{c}{} \\ \hline
GTE-base $*$ & 57.92 & 79.25 & 77.98 & \textbf{98.22} & 68.93 & 46.36 & 42.42 & 71.90 & 73.04 \\
\hspace{1em} +Baseline & \textbf{60.40} & 80.87 & 81.12 & 98.04 & 69.75 & 46.76 & 43.18 & 72.88 & 78.34 \\
\hspace{1em} +Ours (LLR) $\dagger$ & \textbf{60.40} & \textbf{82.25} & \textbf{83.10} & 98.05 & \textbf{70.16} & \textbf{47.23} & \textbf{45.46} & \textbf{75.53} & \textbf{79.12} \\ \hline \hline
Snowflake-Arctic-Embed-l $*$ & 23.76 & 63.07 & 54.44 & 32.55 & 64.78 & 27.75 & 13.64 & 54.28 & 41.30 \\
\hspace{1em} +Baseline & 64.36 & 81.15 & 90.76 & 89.38 & 85.32 & 67.25 & 38.64 & 69.49 & 85.22 \\
\hspace{1em} +Ours (CLR) $\dagger$ & \textbf{72.28} & \textbf{86.49} & \textbf{91.06} & \textbf{91.49} & \textbf{89.87} & \textbf{71.78} & \textbf{40.15} & \textbf{78.86} & \textbf{87.84} \\ \hline
\multicolumn{4}{l}{\small $*$ denotes the original model.}\\ 
\multicolumn{10}{l}{\small All three strategies yield improvements over the baseline, $\dagger$ indicates the strategy achieving the largest gain for this model.}\\
\end{tabular}%
}
\end{table*}


The results in Table~\ref{tab:validate_result} confirm the effectiveness of the Lexical-Semantic Bridge across diverse retrieval tasks and model backbones. Among the bridging strategies, CLR consistently achieves the best performance on all tasks with both DistilBERT and MPNet, particularly improving DistilBERT’s keyword retrieval accuracy from 58.72 to 79.03. For P-o-P tasks, the most significant gains occur on shorter spans (P-o-P 16), with diminishing returns as segment length increases due to richer contextual information. Therefore, subsequent experiments focus on P-o-P-16 to better highlight fine-grained improvements.


Table~\ref{tab:plugintosotamodel} demonstrates the adaptability of the bridge across strong embedding models. All strategies yield performance gains, though the optimal choice depends on the underlying architecture. For GTE, LLR improves fine-grained retrieval while maintaining strong semantic performance. For Snowflake-Arctic-Embed-l, CLR achieves consistent improvements across fine-grained tasks while preserving semantic retrieval. These results show that the Lexical-Semantic Bridge enhances existing models without architectural changes or compromising overall performance.

\begin{figure}[h] 
  \centering
  \includegraphics[width=1.0\linewidth]{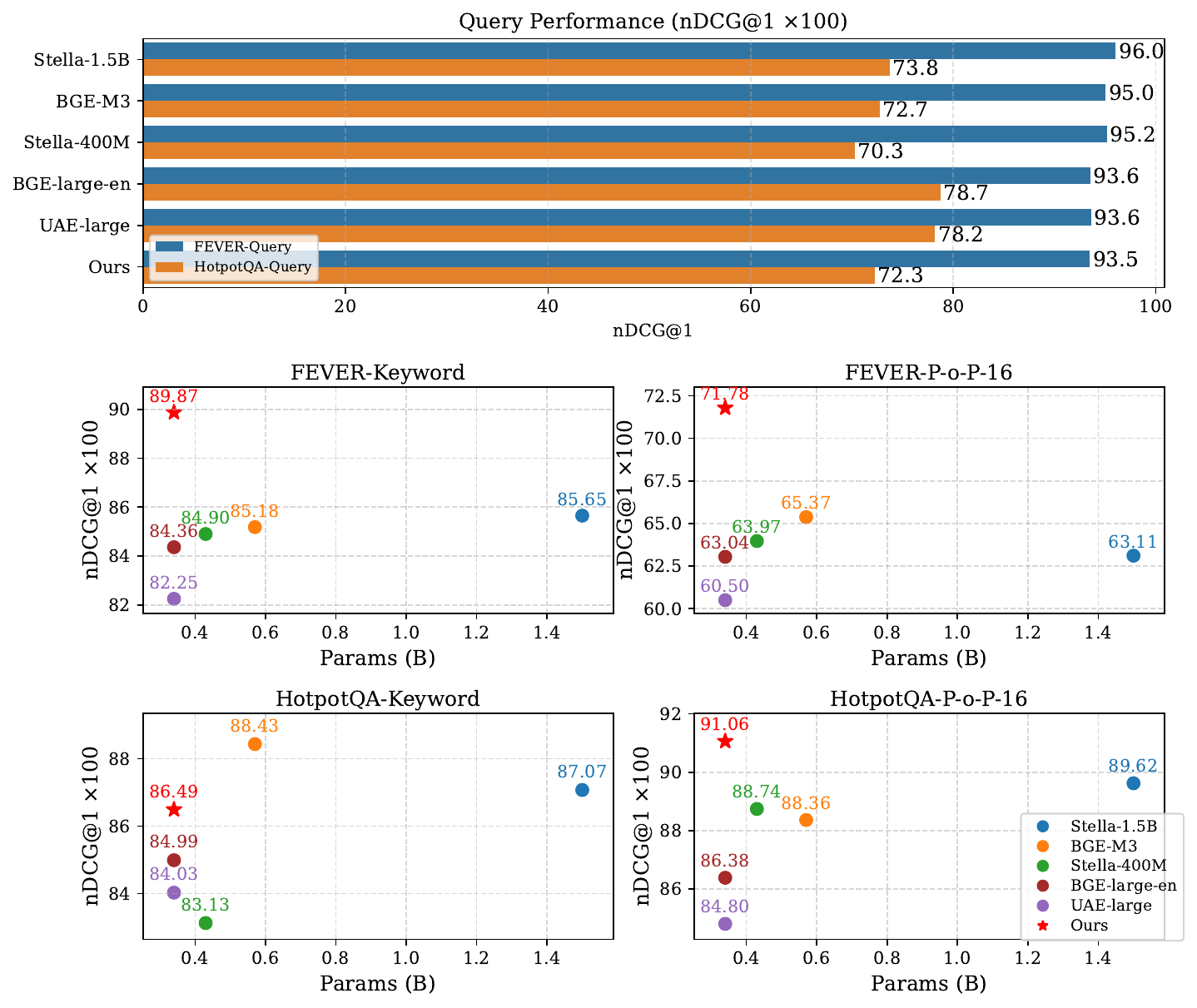}
  \caption{Comparative performance of dense retrieval models. The top bar chart shows nDCG@1~($\times$~100) on semantic queries from HotpotQA and FEVER. The bottom scatter plots illustrate model sizes and performance on fine-grained retrieval tasks.}
  \label{fig:sotamodelresult}
\end{figure}

Figure~\ref{fig:sotamodelresult} compares our method with representative state-of-the-art dense retrieval models, including Stella-en-1.5B-v5, Stella-en-400M-v5~\cite{zhang2024jasper}, BGE-M3~\cite{chen2024bge}, UAE-large~\cite{li2023angle}, and BGE-large-en-v1.5~\cite{xiao2024c}. Our model is based on Snowflake-Arctic-Embed-l and trained with the CLR strategy described in Sec.~\ref{sec:finetune_strategy}. For fairness, BGE-M3 is evaluated using only its dense encoder, excluding sparse and multi-vector components. ColBERT is not included, as its late interaction design deviates from the embedding-based retrieval framework defined in Eq.~\eqref{eq:embedding_eq} and Eq.~\eqref{eq:cosim}. In semantic retrieval, our model achieves competitive nDCG@1 scores on HotpotQA and FEVER with significantly fewer parameters. For fine-grained tasks, it outperforms all models of similar size and remains competitive with larger ones, particularly on P-o-P-16. These results demonstrate that the Lexical-Semantic Bridge improves retrieval quality while maintaining strong parameter efficiency.

\subsection{Adaptation Experiments for Enhancing Image Retrieval Models}
\label{sec:ad_to_image}


To assess the generalizability of the Lexical-Semantic Bridge beyond textual retrieval, we apply it to visual embedding models and evaluate BEiT-base-patch16-224 on two fine-grained image retrieval benchmarks: CUB200 and StanfordCars. As vision models like BEiT tokenize images into discrete patch tokens analogous to textual words~\cite{baobeit}, our enhancement strategies (SLR, LLR, and CLR) can be directly applied without architectural changes.

As shown in Table~\ref{tab:beit_image_retrieval}, all three strategies consistently outperform the baseline, indicating effective transfer to vision transformers. The LLR strategy achieves the highest accuracy on CUB200 (76.96), while CLR performs best on StanfordCars (50.14). These results demonstrate that the proposed method adapts well to non-text modalities and serves as a general enhancement for fine-grained retrieval without requiring architectural modifications.


\begin{table}[ht]
\centering
\caption{Performance of BEiT-base-224 on image retrieval benchmarks (nDCG@1~($\times$~100)).}
\label{tab:beit_image_retrieval}
\begin{tabular}{lcc}
\toprule
\textbf{BEiT-base-224} & \textbf{CUB200} & \textbf{StanfordCars} \\
\midrule
Baseline & 55.28 & 37.82 \\
SLR & 66.02 & 40.61 \\
LLR & \textbf{76.96} & 43.53 \\
CLR & 76.55 & \textbf{50.14} \\
\bottomrule
\end{tabular}
\end{table}

\subsection{Comparison with learnable sparse representation models}
\label{sec:com_lsr}

Learnable sparse representations serve as an enhanced form of traditional sparse retrieval, inherently well-suited for fine-grained tasks due to their reliance on term-level activations. We compare our method with two representative LSR approaches: SPLADE and SPARTA, using their publicly released best-performing checkpoints (SPLADE v2-max and sparta-msmarco-distilbert-base-v1). For both models, we follow their original configuration and compute similarity using dot product over the sparse representations. Experimental results are shown in Table~\ref{tab:compare_lsr}. Our method outperforms both SPLADE and SPARTA in the Query setting on HotpotQA and FEVER, indicating stronger performance in semantic retrieval. SPLADE achieves competitive results in Keyword and P-o-P tasks, as expected given its design for fine-grained lexical matching. In contrast, SPARTA performs consistently worse across all tasks, suggesting limited adaptability in complex retrieval scenarios.

\begin{table}[h]
\centering
\caption{Comparison with sparse retrieval models on HotpotQA and FEVER (nDCG@1$\times$100). The best performance is marked in \textbf{bold}, and the second-best performance is \uline{underlined}.}
\label{tab:compare_lsr}
\resizebox{0.9\columnwidth}{!}{%
\setlength{\tabcolsep}{3pt}
\renewcommand{\arraystretch}{1.1}
\small
\begin{tabular}{l|ccc|ccc}
\hline
\multirow{2}{*}{Model} & \multicolumn{3}{c|}{\textbf{HotpotQA}} & \multicolumn{3}{c}{\textbf{FEVER}} \\
 & Query & Keyword & P-o-P-16 & Query & Keyword & P-o-P-16 \\
\hline
SPLADE & \underline{67.82} & \textbf{87.85} & \textbf{93.66} & \underline{93.35} & \textbf{91.29} & \textbf{77.45} \\
SPARTA & 56.93 & 76.92 & 74.50 & 87.62 & 83.82 & 58.24 \\
Ours & \textbf{72.28} & \underline{86.49} & \underline{91.06} & \textbf{93.49} & \underline{89.87} & \underline{71.78} \\
\hline
\end{tabular}
}
\end{table}

\subsection{Ablation Study for Effectiveness of Query Encoder Enhancement}
\label{sec:ablation_qencoer}


To verify the effectiveness of integrating the Lexical-Semantic Bridge into the query encoder, we conduct ablation experiments comparing three configurations: (1) applying the bridge to the query encoder only, (2) to the passage encoder only, and (3) to both encoders. Table~\ref{tab:ablation_encoder} reports the results based on DistilBERT with CLR across multiple tasks and datasets.


The results show that applying lexical enhancement only to the query encoder consistently achieves the best performance, while passage-side enhancement leads to noticeable drops, possibly due to the already high information density in passages, which may introduce redundancy or amplify irrelevant signals. Enhancing both encoders simultaneously brings no further gain and may even reduce effectiveness due to the increased complexity of aligning two modified representations. These findings confirm that the query, as the main information-seeking component, benefits most from lexical enhancement. This enhancement strategy aligns with the core objective of retrieval tasks and helps close the lexical-semantic gap without disrupting the semantic integrity of passage representations. This supports our framework design, as illustrated in Figure~\ref{LexSem_bridge}.


\begin{table}[h]
\centering
\caption{Ablation results of Lexical-Semantic Bridge encoder head variants (nDCG@1 $\times$ 100) on DistilBERT with CLR strategy}
\label{tab:ablation_encoder}
\resizebox{0.9\columnwidth}{!}{%
\begin{tabular}{llccc}
\toprule
\textbf{Dataset} & \textbf{Task} & \textbf{Query Only} & \textbf{Passage Only} & \textbf{Both Heads} \\
\midrule
\multirow{3}{*}{HotpotQA} 
    & Query     & \textbf{44.06} & 38.61 & 39.11 \\
    & Keyword   & \textbf{79.03} & 75.07 & 77.61 \\
    & P-o-P 16  & \textbf{86.20} & 83.48 & 85.20 \\
\midrule
\multirow{3}{*}{FEVER} 
    & Query     & \textbf{68.51} & 54.47 & 62.32 \\
    & Keyword   & \textbf{72.81} & 62.27 & 67.03 \\
    & P-o-P 16  & \textbf{58.31} & 49.30 & 52.30 \\
\midrule
\multirow{3}{*}{NQ} 
    & Query     & \textbf{28.79} & 24.24 & 23.49 \\
    & Keyword   & \textbf{67.78} & 61.97 & 65.02 \\
    & P-o-P 16  & \textbf{79.20} & 76.22 & 77.66 \\
\bottomrule
\end{tabular}%
}
\end{table}

\subsection{Ablation Study for Effectiveness of Lexical-Semantic Bridge}
\label{exp:ablation_lsb}

\begin{figure}[h] 
  \centering
  \includegraphics[width=0.6\linewidth]{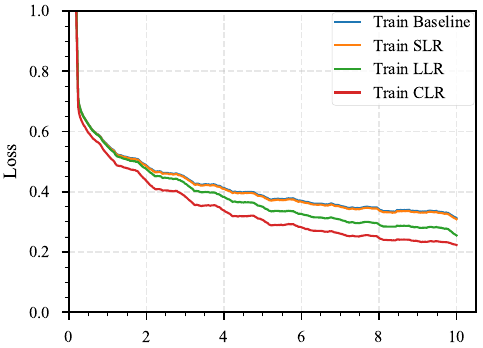}
  \caption{Training Loss with Lexical-Semantic Bridge (DistilBERT)}
  \label{TrainLossCurve}
\end{figure}

Figure~\ref{TrainLossCurve} shows the training dynamics of different variants of the lexical-semantic bridge using DistilBERT. SLR maintains similar convergence to the baseline while effectively integrating term frequency features, confirming that statistical lexical information can be incorporated without hindering semantic learning. LLR and CLR yield modest improvements in loss reduction, suggesting that learned lexical mappings and contextual information benefit optimization. CLR achieves the lowest loss, indicating that leveraging pre-trained MLM heads enables better modeling of nuanced lexical-semantic relationships.


To further assess the intrinsic capabilities of each lexical variant, we conduct ablation experiments by removing fusion with dense vectors ($Q_{dense}$). As shown in Table~\ref{tab:sparse_linear_exp}, CLR performs well across tasks even without dense fusion, highlighting its ability to capture both semantic and lexical signals. While SLR and LLR are less effective for semantic search, they improve performance in fine-grained tasks, indicating their strength in modeling specific lexical patterns. This task-dependent behavior supports our theoretical analysis in Sec.~\ref{sec:method_detail}, where semantic preservation and fine-grained sensitivity are discussed. Overall, these results show that the Lexical-Semantic Bridge, especially CLR, enhances representations by combining contextual understanding with lexical information, helping to bridge the lexical-semantic gap.

\begin{table}[h]
\centering
\caption{nDCG@1 ($\times$100) of Lexical-Only Paradigms Without Dense Fusion}
\label{tab:sparse_linear_exp}
\resizebox{0.65\columnwidth}{!}{%
\begin{tabular}{llccc}
\toprule
\textbf{Dataset} & \textbf{Task} & \textbf{SLR} & \textbf{LLR} & \textbf{CLR} \\
\midrule
\multirow{3}{*}{HotpotQA} 
  & Query     & 0.99  & 8.42  & \textbf{47.03} \\
  & Keyword   & 4.84  & 34.60 & \textbf{76.37} \\
  & P-o-P 16  & 31.68 & 50.44 & \textbf{81.68} \\
\midrule
\multirow{3}{*}{FEVER}    
  & Query     & 7.55  & 16.41 & \textbf{76.99} \\
  & Keyword   & 6.59  & 27.87 & \textbf{71.45} \\
  & P-o-P 16  & 13.94 & 26.75 & \textbf{54.04} \\
\midrule
\multirow{3}{*}{NQ}       
  & Query     & 0.76  & 6.82  & \textbf{26.52} \\
  & Keyword   & 3.57  & 26.70 & \textbf{65.60} \\
  & P-o-P 16  & 26.14 & 45.36 & \textbf{73.16} \\
\bottomrule
\end{tabular}%
}
\end{table}

\section{Conclusion}
\label{sec:conclusion}

In this work, we propose the Lexical-Semantic Bridge, a modular framework designed to enhance dense representation by augmenting query representations with fine-grained lexical information while preserving semantic expressiveness. Our method introduces three complementary paradigms, namely Statistical Lexical Representation (SLR), Learned Lexical Representation (LLR), and Contextual Lexical Representation (CLR), to extract and integrate lexical signals from dense encoders, and can be readily adapted to both text and image retrieval models. Extensive experiments on fine-grained and standard retrieval tasks demonstrate that our framework consistently improves retrieval effectiveness across multiple datasets and architectures, offering a flexible and efficient enhancement to existing dense retrievers.



\begin{ack}
This research is supported by National Natural Science Foundation of China (Grant No.62276154);
Research Center for Computer Network (Shenzhen) Ministry of Education, the Natural Science Foundation of Guangdong Province (Grant No.2023A1515012914 and 440300241033100801770);
Basic Research Fund of Shenzhen City (Grant No.JCYJ20210324120012033, JCYJ20240813112009013 and GJHZ20240218113603006);
The Major Key Project of PCL for Experiments and Applications (PCL2024A08).
\end{ack}



\bibliography{mybibfile}

\end{document}